\documentclass[sigconf]{acmart}

\AtBeginDocument{%
  }

\microtypesetup{expansion=false}
\usepackage{booktabs}
\usepackage{multirow}
\usepackage{xcolor}
\usepackage{graphicx}
\usepackage{comment}
\usepackage{enumitem}

\setcopyright{acmlicensed}
\copyrightyear{2027}
\acmYear{2027}
\acmDOI{XXXXXXX.XXXXXXX}
\acmISBN{978-1-4503-XXXX-X/27/08}

\begin{document}

\title{LLM-Based Generative Retrieval for Snapchat Content Recommendation}

\author{Liam Collins$^*$, Jiwen Ren$^*$, Donald Loveland, Bhuvesh Kumar, Clark Mingxuan Ju, Xuan Guo, Mo Li, Alvin Hou, Yi Cui, Peng Yang, Jian Wang, Saud Afzal Shafi, Nga Than, Ruiming Lu, Wenfeng Zhuo, Dongheng Li, Lili Zhang, Mingtao Zhang, Jinchao Ye, Jiacong He, Vincent Xue, Chunhui Zhu, Neil Shah}
\affiliation{%
  \institution{Snap, Inc.}
  \country{USA}
}
\email{\{lcollins2, jren2, nshah\}@snapchat.com}


\renewcommand{\shortauthors}{Collins, Ren, et al.}

\renewcommand{\shortauthors}{[Authors]}

\begin{abstract}
Pretrained large language models (LLMs) are promising retrieval engines because they combine rich semantic priors, strong sequence modeling capabilities, and favorable scaling behavior. However, turning a pretrained LLM into a generative retriever in production deployment raises several challenges: the model must learn an internal item vocabulary that was absent from pretraining, and generate valid item identifiers under strict latency and cost constraints. 
We address these challenges through the design and launch of \\SnapLGR, an LLM-based generative retrieval system for  short-video recommendation at Snapchat. The system is built around three main designs. 
First, we construct semantic identifiers (SIDs) from multimodal item embeddings and enhance them with Personalized PageRank (PPR)-based co-engagement contrastive learning, resulting in improved codebook utilization, reduced collisions, and infused collaborative signal. 
Second, we use continued pretraining (CPT) to ground the introduced SID tokens before supervised fine-tuning (SFT) on user interaction sequences. 
Third, we make SnapLGR serving practical through TensorRT-LLM CUDA-backed beam search and a decentralized worker-loop architecture.
In a live A/B test, the launched system increased View Time by 0.37\%, Time Spent by 0.09\%, Deep Sessions by 0.18\%, and Deep Sessions Unique User by 0.11\% relative to the existing TIGER-style generative retrieval baseline. {We then decompose this offline gap under a fixed tokenizer and quantify the gains due to model architecture, scaling, and pretraining.} Overall, our deployment shows that successful production SnapLGR requires joint design across representation learning, vocabulary grounding, and efficient training and serving.
\end{abstract}



\maketitle

\begingroup
\renewcommand\thefootnote{}
\footnotetext{$^*$Indicates equal contribution.}
\endgroup

\section{Introduction}

Large-scale recommendation systems rely heavily on retrieval to efficiently filter massive item corpora for downstream ranking. Generative retrieval (GR) reformulates this task as sequence generation: given a user's interaction history, a model generates semantic IDs (SIDs)—discrete token sequences capturing item semantics—for the next items the user is likely to engage with. Following TIGER~\cite{rajput2023recommender}, initial GR systems have relied on small, randomly initialized encoder-decoder transformers such as T5~\cite{raffel2020exploring} as the backbone SID generation model. 
Nevertheless, replacing these backbones with pretrained decoder-only large language models (LLMs) offers immense promise. LLMs bring rich semantic priors, advanced sequence modeling capabilities, and predictable scaling~\cite{kaplan2020scaling,hoffmann2022training}, driving significant industrial interest in LLM-based GR~\cite{he2026plum,zhai2024actions,deng2025onerec,zhou2025onerectr,firooz2025360brew,han2025mtgr,yi2025recgpt,yang2025cobra,huang2025towards}.

However, transitioning an LLM into an active generative retriever for a large-scale production system introduces coupled modeling and infrastructure challenges: 
\begin{enumerate}[leftmargin=*]
  \item \textbf{Tokenizer Design:} SIDs are the interface between the retriever and the item corpus, thus are critical to GR performance \cite{ju2026semanticids}.
  The challenge lies in designing a tokenizer that effectively compresses semantic features, and potentially collaborative signals as well, into SIDs, while avoiding severe token collisions. 
  \item \textbf{Vocabulary Grounding:} SIDs do not exist in the LLM's pretraining vocabulary. The model must semantically ground these new tokens into its existing internal knowledge space before it can accurately leverage its pretrained semantic understanding for recommendation.
  \item \textbf{High-Throughput Training and Serving:} The system must regularly train and execute wide-beam inference at large scale all within a strict GPU budget in spite of LLMs' pronounced computational complexity.
\end{enumerate}

Despite growing interest in LLM-based GR, it is still not well-understood how to jointly tackle these deployment challenges. Moreover, despite the promise of LLM-based GR, its performance gains over legacy TIGER-style GR systems at production scale remain unclear. 
This raises two critical questions: (1) How should we build an LLM-based generative retrieval system to overcome the aforementioned production challenges? and (2) What are the tangible online business gains  and the underlying sources of offline performance improvements of this system over legacy TIGER-style baselines?

We address these questions through the design, production launch, and empirical analysis of \textbf{SnapLGR}: \textbf{Snap}chat \textbf{L}anguage-based \textbf{G}enerative \textbf{R}etrieval,  an end-to-end LLM-based GR system for  short video recommendation at Snapchat, a widely-used social media platform. Our main contributions are:
\begin{itemize}[leftmargin=*]
  \item \textbf{Multimodal, Behavioral SIDs:} We build SIDs by residually quantizing Qwen3-VL~\cite{li2026qwen3vlembedding} multimodal embeddings. We further introduce a Personalized PageRank (PPR)~\cite{haveliwala2002topic} co-engagement contrastive loss that pulls co-engaged items together in the SID space, infusing collaborative signal and reducing collisions.
  \item \textbf{Two-Stage Vocabulary Grounding:} We achieve semantic grounding by decoupling vocabulary learning from task tuning. A frozen-LLM Continued Pretraining (CPT) stage first anchors new SID tokens in textual semantics, followed by Supervised Fine-Tuning (SFT) on user interaction histories.
  \item \textbf{High-Throughput Training and Inference:} We co-optimize our infrastructure to support frequent periodic retraining and massive-scale batch inference. For training, we achieve a $3.63\times$ throughput increase via tuned sequence packing, FlashAttention-2~\cite{dao2023flashattention2} variable-length kernels, and graph compilation. For inference, we eliminate Python framework bottlenecks using TensorRT-LLM's~\cite{trtllm}  CUDA-backed beam search, a decentralized worker-loop architecture, and asynchronous I/O, delivering roughly $45.7\times$ speedup in per-GPU throughput over our legacy baseline.
\item \textbf{Production Launch:} In a 7-day live A/B test, our launched system achieved a +0.37\% increase in View Time, a +0.09\% increase in Time Spent, a +0.18\% increase in Deep Sessions, and a +0.11\% increase in Deep Sessions Unique User over the incumbent TIGER-style GR baseline. 
\item \textbf{Offline Analysis:} We conduct a series of offline studies to understand the reasons for SnapLGR's improvement over the legacy TIGER-style system, finding that the decoder-only backbone and model scaling contribute most significantly to the gain.
\end{itemize}

SnapLGR is currently deployed in production as a retrieval source for a short-form video retrieval platform at Snapchat. Overall, our successful launch demonstrates that industrial-scale generative retrieval is an end-to-end co-design problem, requiring close alignment across item representation, vocabulary grounding, and training and inference engineering.

\section{System Design}
\label{sec:system}

SnapLGR comprises four interdependent components: a video tokenizer that encodes videos as SIDs  (\S\ref{sec:sids}); a two-stage LLM training pipeline (\S\ref{sec:training}); an inference pipeline including SID-to-video materialization, which connects generation to the serving stack (\S\ref{sec:integration}); and high-throughput training and inference systems (\S\ref{sec:inference}).  
We provide additional implementation details in Appendix~\ref{app:pipeline-details}.

\begin{figure*}[t]
  \centering
  \includegraphics[width=0.92\textwidth]{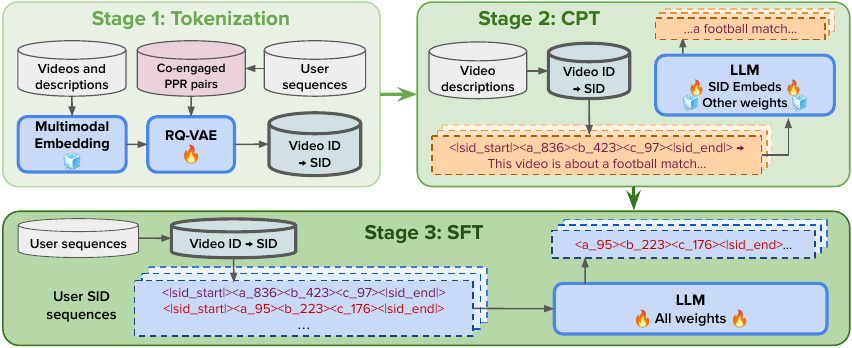}
  \caption{SnapLGR Training pipeline. \emph{Stage 1: Tokenization} (\S\ref{sec:sids}). We train an RQ-VAE on multimodal Qwen3-VL embeddings with PPR-based co-engagement supervision to produce the video-to-SID map. \emph{Stage 2: CPT} (\S\ref{sec:cpt}). We ground newly added SID embeddings in the LLM's vocabulary using video text descriptions. \emph{Stage 3: SFT} (\S\ref{sec:sft}). We fine-tune the full LLM to predict future engaged SIDs from users' SID interaction histories.}
  \label{fig:system}
\end{figure*}

\subsection{Multimodal, Behavior-aware SIDs}
\label{sec:sids}

Semantic IDs (SIDs) are the canonical item representation in GR. SIDs assign multiple tokens per item, which allows for exponentially reducing embedding table dimensions, caters to the sequence modeling and generation capabilities of GR backbones, and enables encoding cross-item relationships via shared tokens.
Formally, the SID for items $i \in \mathcal{I}$ has $L$ levels. Its code at level $\ell$ is
$z_i^\ell\in\{1,\ldots,K_\ell\}$, where $K_\ell$ is the size of the corresponding codebook; we denote the complete SID by $\mathbf z_i=(z_i^1,\dots,z_i^L)$.

An effective tokenizer for production video retrieval must map items to discrete tokens in a way that captures both rich multimodal semantics and collaborative user engagement signals, allowing the generative backbone to effectively learn and leverage cross-video relationships. At the same time, the tokenizer must be heavily collision-resistant, dispersing video assignments broadly across the code space to prevent large clusters of distinct videos from collapsing into identical SIDs. This ensures the generated codes maintain fine representation granularity and simplifies downstream SID-to-video materialization over a large-scale corpus. To achieve this, our construction combines a multimodal embedding, residual quantization, and graph-derived co-engagement supervision.

\noindent\textbf{Multimodal embedding.} Let $v_i$ and $t_i$ denote the video and text description for video $i$, where the text description is generated by Gemini 2.5 Flash~\cite{comanici2025gemini}. To capture a strong baseline of item semantics, we use a powerful multimodal embedding model, Qwen3-VL-Embedding-8B~\cite{li2026qwen3vlembedding}, to map these features to a joint representation
$\mathbf{e}_i =\boldsymbol\phi(v_i,t_i)$.

\noindent\textbf{Residual quantization.} After embedding each video, we train a residual quantizer to map these embeddings to SIDs, following the RQ-VAE design of \citet{ju2026semanticids}. An encoder $E$ maps the normalized embedding to an initial residual of width $d_q$,
$\mathbf r_i^1=E(\operatorname{LN}(\mathbf e_i))$, where we apply LayerNorm (LN) to stabilize training and prevent codebook collapse. Each level has a codebook
$\mathbf C_\ell\in\mathbb R^{K_\ell\times d_q}$. At level $\ell$, we compute cosine similarity scores $s_{ic}^{\ell} =\text{cossim}{(\mathbf r_i^\ell, \mathbf C_\ell[c])}$ against every code $c \in \{1,\dots,K_\ell\}$, then subtract the code with the max score:
\begin{equation}
  \mathbf r_i^{\ell+1} =\mathbf r_i^\ell-\mathbf C_\ell[z_i^\ell], \quad \text{where}\quad z_i^\ell =\arg\max_{c\in\{1,\ldots,K_\ell\}}s_{ic}^{\ell}.
  \label{eq:rq_assignment}
\end{equation}
Thus the first code approximates the encoder output, and later codes successively encode what earlier levels did not capture.

We use the hard-forward, full-codebook straight-through estimator from \citet{ju2026semanticids} to pass gradients through the discrete $\arg\max$ in Equation~\eqref{eq:rq_assignment}. Let
$\mathbf s_i^\ell=(s_{i1}^\ell,\ldots,s_{iK_\ell}^\ell)$, let $\mathbf a_i^\ell=\operatorname{softmax}(\mathbf s_i^\ell)$, and let $\operatorname{sg}(\cdot)$ denote the stop gradient operation. During training, the selected vector is represented as:
\begin{equation}
  \widetilde{\mathbf q}_i^\ell=
  \mathbf C_\ell[z_i^\ell]
  +(\mathbf a_i^\ell)^\top\mathbf C_\ell
  -\operatorname{sg}\!\left((\mathbf a_i^\ell)^\top\mathbf C_\ell\right).
  \label{eq:full_codebook_ste}
\end{equation}
The last two terms cancel in the forward pass, so the SID remains a hard assignment. In the backward pass, however, they provide gradients through the similarity scores and all codebook rows, reducing the risk that unselected codes remain permanently unused. The decoder reconstructs the normalized input as
$\widehat{\mathbf x}_i=D(\operatorname{LN}(\sum_{\ell=1}^L\widetilde{\mathbf q}_i^\ell))$.
For a minibatch of $N$ video embeddings, the RQ-VAE loss consists of reconstruction and commitment losses:
\begin{align}
  \mathcal L_{\mathrm{rec}}
  &=\frac{1}{Nd}\sum_{i=1}^N
    \left\lVert\operatorname{LN}(\mathbf e_i)-\widehat{\mathbf x}_i\right\rVert_2^2, \\
  \mathcal L_{\mathrm{com}}
  &=\frac{1}{2NLd_q}\sum_{i=1}^N\sum_{\ell=1}^L
    \left(
      \left\lVert\mathbf r_i^\ell-\operatorname{sg}(\mathbf q_i^\ell)\right\rVert_2^2
      +\left\lVert\operatorname{sg}(\mathbf r_i^\ell)-\mathbf q_i^\ell\right\rVert_2^2
    \right),
  \label{eq:rq_losses}
\end{align}
where $\mathbf q_i^\ell=\mathbf C_\ell[z_i^\ell]$. The two commitment terms pull encoder residuals toward their assigned codes and update code vectors toward assigned residuals, respectively.

\noindent\textbf{Co-engagement contrastive loss.} Following PLUM~\cite{he2026plum}, we additionally inject collaborative signal into the SIDs via a co-engagement-based contrastive loss to improve their behavioral alignment and recommendation utility. We implement this by first feeding the video's sum of quantized latents through a projection head $G_c$:
\begin{equation}
  \mathbf g_i=\operatorname{normalize}\!\left(
  G_c\!\left(\sum_{\ell=1}^L\widetilde{\mathbf q}_i^\ell\right)
  \right).
  \label{eq:sid_projection}
\end{equation}
For a minibatch of co-engaged video pairs $\{(i_p,i_p^+)\}_{p=1}^B$, we use $i_p^+$ as the positive for anchor $i_p$ and the other $B-1$ positive-side videos as in-batch negatives:
\begin{equation}
  \mathcal L_{\mathrm{co}}=-\frac{1}{B}\sum_{p=1}^B
  \log\frac{\exp(\mathbf g_{i_p}^\top\mathbf g_{i_p^+}/\tau)}
  {\sum_{p'=1}^B\exp(\mathbf g_{i_p}^\top\mathbf g_{i_{p'}^+}/\tau)}.
  \label{eq:sid_contrastive}
\end{equation}
Here $\tau>0$ is the contrastive temperature. The complete tokenizer objective is
$\mathcal L=\mathcal L_{\mathrm{rec}}+\lambda_{\mathrm{com}}\mathcal L_{\mathrm{com}}+\lambda_{\mathrm{co}}\mathcal L_{\mathrm{co}}$. 

\noindent\textbf{PPR positive-pair construction.} To construct the closely related positive video pairs $\{(i_p,i_p^+)\}$ required for Equation~\eqref{eq:sid_contrastive}, we build an undirected bipartite user--video graph $\mathcal G=(\mathcal U\cup\mathcal V,\mathcal E)$ from eligible engagements using an internal large-scale graph processing framework. We compute Personalized PageRank (PPR)~\cite{haveliwala2002topic} scores over the two-hop neighborhood for each anchor video $i$ (detailed in Appendix~\ref{app:ppr_details}). Because the graph is bipartite, a two-step random walk connects the anchor to another video $j$ through a shared user, naturally accounting for multiple overlapping interaction paths while avoiding bias toward globally popular nodes. We rank video nodes $j\ne i$ by their PPR score and retain high-scoring $(i,j)$ pairs as contrastive positives. In production, we use the P75 score across all pairs as the minimum threshold for selection.

\subsection{LLM Training Pipeline}
\label{sec:training}

After tokenizing the videos, we need to teach the backbone LLM to recommend over the SID vocabulary. We do this in two stages: continued pretraining (CPT) and supervised fine-tuning (SFT). We use Qwen3-0.6B \cite{qwen3_0.6b_model} as the backbone LLM.

\subsubsection{Continued Pretraining (CPT)}\label{sec:cpt}
The first step of CPT is to append the new SID tokens to the LLM's vocabulary. We randomly initialize their embeddings from a Gaussian distribution with mean and covariance equal to those of the existing embeddings. 
Then, we align the SID embeddings by training them on a video SID-to-description task. In particular, for each video $i$, we get its SID $\mathbf{z}_i$ and text description $t_i$, prompt the LLM to generate $t_i$ given $\mathbf{z}_i$, and compute the loss over the generated text description tokens.
We train only the SID embeddings on this loss while keeping the rest of the model frozen, serving to ground the embeddings in the LLM's existing world knowledge. We refer the reader to Appendix \ref{app:cpt} for an illustrative prompt.

\subsubsection{Supervised Fine-Tuning (SFT)}
\label{sec:sft}
SFT fine-tunes the LLM checkpoint produced by CPT on users' SID interaction sequences.
We use the ChatML prompt format where the user prompt contains the chronological SID history and the assistant prompt has future events, and we compute the loss only on the assistant tokens. 

\subsubsection{Training Lifecycle}
CPT and SFT operate on distinct schedules in our production pipeline. CPT is executed once during the initial base model training phase over a broad corpus of video descriptions to firmly establish the new SID vocabulary. Following this, SFT transitions into hourly incremental training on fresh interaction sequences to maintain temporal relevance.

\subsection{Inference Pipeline}
\label{sec:integration}
\begin{figure}[t]
  \centering
  \includegraphics[width=\columnwidth]{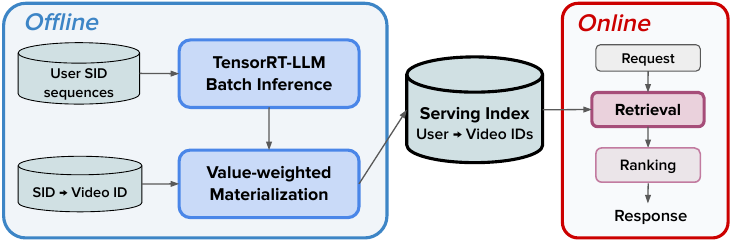}
  \caption{Serving pipeline. \textit{Offline:} daily batch inference generates each user's top-$k$ SIDs. Value-weighted materialization maps the SIDs to candidate videos and writes user--video lists to an {index.} \textit{Online:} given a request, the serving index serves the video ID list as retrieval candidates.}
  \label{fig:inference}
\end{figure}

Figure~\ref{fig:inference} illustrates our end-to-end serving pipeline, which is structurally divided into offline batch generation and online serving stages. In the offline stage, historical user SID sequences are   daily by a TensorRT-LLM batch inference engine to generate predicted future SIDs. Because SIDs represent semantic clusters rather than exact items, they pass through a value-weighted materialization step using a precomputed SID-to-Video ID mapping. The resulting user-to-video candidate lists are written to a low-latency serving index. During the online stage, a user request triggers the retrieval component to fetch these pre-materialized video candidates directly from the serving index. Finally, these candidates are passed downstream to the ranking layer to assemble the final response.

\subsection{Throughput Optimizations}
\label{sec:inference}
Incremental SFT retraining and batch inference over hundreds of millions of sequences requires tailored optimization under strict GPU budgets. Standard general-purpose LLM infrastructure is poorly suited for generative retrieval: training suffers from sequence padding inefficiencies, while inference is bottlenecked by Python runtime overheads during short-sequence wide-beam decoding. We make several optimizations to our training and serving pipelines to maximize hardware utilization and throughput.

\noindent\textbf{Training.} We apply a cumulative series of software optimizations:
\begin{enumerate}[leftmargin=*]
  \item \emph{\texttt{torch.compile}:} We leverage graph compilation to reduce Python framework overhead and fuse basic GPU operations.
  \item \emph{FlashAttention-2 Variable-Length Support:} We utilize FlashAttention-2's built-in variable-length capabilities to compute attention directly over unpadded sequences.
  \item \emph{Dynamic Sequence Packing:} We concatenate user interaction histories into contiguous memory buffers, eliminating wasted computation on padding tokens.
\end{enumerate}
As we discuss in \S\ref{sec:inference_exps}, this optimization stack delivers a $3.63\times$ end-to-end throughput increase over the unoptimized baseline when combined with a hardware migration from NVIDIA A100 to NVIDIA H100 GPUs.

\noindent\textbf{Inference.}
Our serving infrastructure is built around three core designs aimed at achieving high-throughput SID generation:
\begin{enumerate}[leftmargin=*]
  \item \emph{CUDA-Backed Beam Search:} We execute generation using TensorRT-LLM's~\cite{trtllm} CUDA-backed runtime. This ensures beam expansion, scoring, and KV-cache management occur entirely on-GPU, avoiding redundant device-to-host memory copies and minimizing Python scheduling overhead.
  \item \emph{Decentralized Worker-loop Architecture:} We employ a decentralized worker-loop processing model. Each worker independently claims, loads, and processes its assigned Parquet data shards, fully removing centralized driver-node coordination and fan-out bottlenecks from the critical path.
  \item \emph{Asynchronous I/O Overlapping:} To prevent data ingestion and output serialization from stalling GPU execution, we implement background row-group prefetching and isolate Parquet writes to a dedicated asynchronous thread pool. Additionally, we strictly gate non-essential CPU operations, such as unused metric encoding, out of the generation loop.
\end{enumerate}
Ultimately, compounding these architectural changes and I/O overlapping strategies yields end-to-end per-GPU throughput gains of $45.7\times$ over the legacy un-optimized infrastructure. We detail the quantitative impact of each specific optimization in \S\ref{sec:inference_exps}.

\section{SID Quality and CPT Grounding Evaluation}

We begin by evaluating the pre-SFT components of our pipeline offline. In \S\ref{sec:sid_quality_results}, we evaluate the tokenizer's representation quality with a focus on \emph{collision resistance}. In \S\ref{sec:sid_grounding_results}, we evaluate how effectively continued pretraining (CPT) grounds the newly introduced SID tokens into the LLM's vocabulary.

\subsection{SID Quality \& Collision Resistance}
\label{sec:sid_quality_results}

An unbalanced code mapping results in severe token collisions, where disproportionately large sets of stories map to identical SIDs. This degrades the LLM's ability to distinguish subtle semantic differences between items and complicates downstream candidate materialization. 

\begin{figure}[t]
  \centering
  \includegraphics[width=0.88\columnwidth]{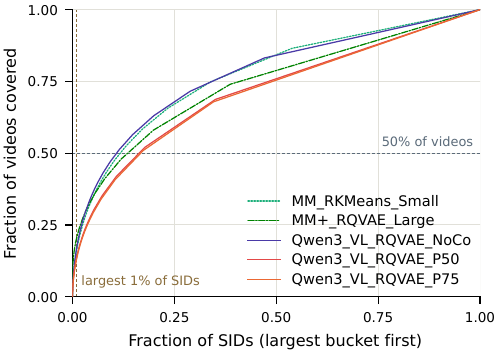}
  \caption{SID collision concentration. SIDs are ordered by decreasing bucket size. Lower curves indicate a more balanced assignment; dashed guides mark the largest 1\% of SIDs and 50\% video coverage.}
  \label{fig:tail_concentration}
\end{figure}

\noindent\textbf{Metrics.} We evaluate collision resistance using four metrics: \emph{utilization} (the proportion of the theoretical code space occupied), \emph{uniqueness} (the fraction of items assigned a unique SID), \emph{top-1\% coverage} (the fraction of stories falling into the largest 1\% of SID buckets), and the \emph{Gini coefficient} of the collision distribution (summarizing the concentration curve in Figure~\ref{fig:tail_concentration}). Lower Gini and top-1\% coverage values indicate that stories are more evenly dispersed across SIDs. Formal metric definitions are in Appendix~\ref{app:sid_metrics}.

\noindent\textbf{Baselines.} We compare our production tokenizer that uses Qwen3-VL embeddings and PPR co-engagement supervision above P75, denoted \textbf{Qwen3\_VL\_RQVAE\_P75}, against four baselines evaluated on the same large-scale video cohort: \textbf{\textbf{MM\_RKMeans\_Small}}, which uses internal multimodal embeddings and residual KMeans \cite{ju2025generative}; \textbf{ MM+\_RQVAE\_Large}, a similar internal baseline that concatenates additional multimodal embeddings and uses RQ-VAE; and two Qwen3\_VL\_RQVAE ablations,  \textbf{\_NoCo} without co-engagement, and \textbf{\_P50} with a lower engagement threshold. All versions use codebook widths (1024, 512, 256) besides  {MM\_RKMeans\_Small} which uses (256, 256, 256).

\begin{table}[t]
  \caption{SID assignment statistics.}
  \label{tab:sid_quality}
  \setlength{\tabcolsep}{4pt}
  \begin{tabular}{@{}lrrrrr@{}}
    \toprule
    \textbf{SID} & \textbf{Util.\%} & \textbf{Uniq.\%} & \textbf{Gini} & \textbf{Top-1\%} \\
    \midrule
    MM\_RKMeans\_Small       & 36.3 & 15.8 & 0.548 & 16.0 \\
    MM+\_RQVAE\_Large       & 45.5 & 29.1 & 0.483 & 20.4 \\
    Qwen3\_VL\_RQVAE\_NoCo          & 31.8 & 16.8 & 0.551 & 16.9 \\
    Qwen3\_VL\_RQVAE\_P50           & 48.2 & 31.3 & 0.432 & 13.6 \\
    \midrule
    Qwen3\_VL\_RQVAE\_P75           & \textbf{49.1} & \textbf{32.1} & \textbf{0.424} & \textbf{13.1} \\
    \bottomrule
  \end{tabular}
\end{table}

 \noindent\textbf{Results.} As shown in Table~\ref{tab:sid_quality}, PPR co-engagement supervision dramatically improves code utilization and collision resistance. Within the Qwen3-VL cohort, adding co-engagement at the P50 threshold increases utilization from 31.8\% to 48.2\% and nearly doubles uniqueness (16.8\% to 31.3\%), with P75 filtering providing marginal further gains. Figure~\ref{fig:tail_concentration} illustrates this improved concentration visually, showing that the Qwen3-VL co-engagement variants display curves closer to the diagonal than the baselines.

\subsection{SID Grounding}
\label{sec:sid_grounding_results}

When new SID tokens are appended to the LLM vocabulary, their embeddings are typically initialized as the mean of existing text tokens with minor variance, meaning they lack  alignment with  the LLM's semantic knowledge space, i.e. they are ungrounded. Here we evaluate how well CPT achieves that grounding.

\noindent \textbf{Metrics.} We use \textbf{Text-Grounding RSA} (Representational Similarity Analysis)~\cite{kriegeskorte2008representational} to evaluate SID grounding. RSA quantifies the geometric correlation between the learned SID embeddings and the LLM's text embeddings of the corresponding item descriptions. 

Formally, for a set of $N$ items, let $\mathbf{t}_i$ denote the mean-pooled base text embedding of item $i$'s description, and $\mathbf{s}_{i,\ell}$ denote the learned embedding of item $i$'s $\ell$-th SID token. For each layer $\ell \in [L]$, we construct two $N \times N$ pairwise cosine distance matrices, $D^{\text{text},\ell}$ and $D^{\text{SID},\ell}$, where elements are defined as $D_{i,j} = 1 - \cos(\mathbf{x}_i, \mathbf{x}_j)$. The Text-Grounding RSA is the Spearman rank correlation $\rho$ between the flattened strictly upper-triangular elements of these matrices, averaged over all $L$ SID layers:
\begin{equation}
  \text{RSA} = \frac{1}{L} \sum_{\ell=1}^L \rho \Big( \text{triu}(D^{\text{text},\ell}), \, \text{triu}(D^{\text{SID}, \ell}) \Big).
\end{equation}
A higher RSA indicates that the learned SID space correctly mirrors the semantic topology of the LLM's preexisting textual knowledge.

\begin{table}[t]
  \centering
  \caption{SID Grounding Evaluation. Post-CPT embeddings yield the strongest proxy for text semantic alignment, though this alignment is mostly lost after SFT.}
  \label{tab:grounding}
  \setlength{\tabcolsep}{36pt}
  \begin{tabular}{@{}lcc@{}}
    \toprule
    \textbf{Probe} & \textbf{RSA $\uparrow$} \\
    \midrule
    Random Initialization               & 0.016 \\
    \midrule
    \textbf{CPT}     &  \textbf{0.322} \\
    {CPT+SFT}                    & 0.167 \\
    SFT Only  &  {0.162} \\
    \bottomrule
  \end{tabular}
\end{table}

\noindent\textbf{Baselines \& Training Stages.} To isolate the impact of our vocabulary grounding strategy, we evaluate the Text-Grounding RSA across four distinct configurations:
\begin{itemize}[leftmargin=*]
  \item \textbf{Random Initialization:} The state of the new SID tokens immediately after being appended to the LLM vocabulary. These embeddings are initialized as Gaussian samples with mean equal to the mean of the base text embeddings and small covariance. 
  \item \textbf{CPT:} The embeddings after CPT, which trains the frozen LLM on a semantic grounding task using item text descriptions to establish text structure before downstream behavioral tuning.
\item \textbf{CPT+SFT:} The embeddings after SFT from the CPT checkpoint, where SFT trains only on SID interaction sequences.
  \item \textbf{SFT Only:} The embeddings after SFT from the pretrained Qwen3-0.6B model, without CPT.
\end{itemize}

\noindent\textbf{Results.} As shown in Table~\ref{tab:grounding}, new SIDs start with almost no correlation to their text embeddings at initialization (RSA of 0.016). Applying CPT establishes a strong initial semantic alignment, raising the RSA significantly to 0.322. This confirms that the continued pretraining stage successfully anchors the novel SID tokens within the LLM's pre-existing textual knowledge space. However, this explicit textual grounding is mostly lost after subsequent behavioral tuning. Following SFT on user interaction sequences, the CPT+SFT configuration experiences a stark drop in RSA to 0.167. This pulls it down to a level  similar to training with SFT alone (0.162), which itself achieves a non-trivial RSA likely due to natural correlations between item semantics and user co-engagement patterns. This convergence suggests that while CPT effectively injects initial semantic priors, the sequential recommendation objectives of SFT aggressively reshape the embedding space, overriding pure text semantics to prioritize collaborative relationships. Nevertheless, the strong initial grounding provided by CPT may still confer optimization benefits during SFT training. {We quantify the downstream retrieval gain due to CPT in \S\ref{sec:ablations}.}

\section{Retrieval Evaluation}
\label{sec:experiments}

Having evaluated the results of the tokenization and CPT stages, we now evaluate the results of SFT through retrieval performance. We guide our evaluation by four research questions:
\begin{itemize}[leftmargin=*]
  \item \textbf{RQ1 (Offline Quality):} How does SnapLGR perform against the TIGER-style T5 baseline at offline SID retrieval?
  \item \textbf{RQ2 (Improvement Attribution):} Which design choices drive SnapLGR's offline gains?
  \item \textbf{RQ3 (Training and Inference Efficiency):} How much do our training and inference optimizations enhance throughput?
  \item \textbf{RQ4 (Online Impact):} What is the online production impact of the full SnapLGR system?
\end{itemize}

To our knowledge, we are the first to systematically study the performance gap between LLM-based GR and the TIGER-style T5 baseline at industrial scale. Moreover, our online results demonstrate the practical utility of SnapLGR. We provide additional offline studies, including studying the effects of training data volume and beam width, in Appendix \ref{app:additional_results}.

\subsection{Offline Training and Evaluation Setup}

\noindent\textbf{Data.} SFT uses a 1-day window, and evaluation uses the immediately following hour. CPT uses a 1-month window of video data that ends 1 week prior to the SFT training date. Evaluation uses up to 10 future items per request as ground-truth labels. 

\noindent\textbf{Metrics.} Unless otherwise noted, offline experiments evaluate SID-level generation before materialization. We use Recall@$k$ (abbreviated R@$k$) and  Pass@$k$ (P@$k$), and also report Pos-1-Pass@$k$ in extended results in Appendix \ref{app:additional_results} (see also for full metric definitions).
We use beam search with width 32, and Recall@32 is our primary offline metric to match the production beam width of 32.

\noindent\textbf{TIGER-style Baseline (T5).} Our legacy generative retrieval baseline follows the TIGER~\cite{rajput2023recommender} paradigm, utilizing a 13M-parameter T5 encoder--decoder architecture initialized from scratch. {The deployed production control uses CLIP-based SIDs with coarser codebook widths, $(256,256,256)$ rather than our $(1024,512,256)$.} To cleanly isolate model architecture and capacity, for our offline SID retrieval comparisons we train this baseline using the exact same Qwen3-VL SIDs, training data, and evaluation protocol as our LLM models. We refer to this baseline simply as \textbf{T5} hereafter.

\subsection{Overall Model Comparison (RQ1)} To address \textbf{RQ1}, Table~\ref{tab:t5_vs_llm_condensed} compares the generative retrieval performance of our SnapLGR architecture against the legacy T5 encoder-decoder baseline. By fully harnessing LLM pretraining, continued pretraining (CPT), and supervised fine-tuning (SFT), SnapLGR dramatically outperforms the legacy system across all key top-$k$ retrieval metrics. Most notably, SnapLGR achieves a $2.51\times$ increase in Pass@10 (21.78\% vs.\ 8.685\%) and a $2.27\times$ increase in Pass@32. Similar magnitude improvements are observed in item recall, with Recall@32 reaching 11.11\% compared to the baseline's 4.624\% ($2.40\times$ lift). These results underscore the substantial representational and ranking capacity advantages of the pretrained causal LLM architecture over traditional encoder-decoder setups for semantic identifier retrieval. 

\subsection{Attribution of Improvement Sources (RQ2)} \label{sec:ablations}
 We evaluate {three} non-mutually-exclusive hypotheses to understand \emph{why} SnapLGR outperforms legacy GR baselines. {Because this offline comparison holds the tokenizer fixed, we ablate the tokenizer separately, as a design choice rather than a driver of the gap.}

\begin{table}[t]
  \centering
  \caption{Offline SID retrieval performance comparison over Qwen3-VL SIDs. SnapLGR yields a greater than $2\times$ improvement across all top-$k$ metrics relative to the  T5 baseline.}
  \label{tab:t5_vs_llm_condensed}
  \setlength{\tabcolsep}{7pt}
  \begin{tabular}{@{}lrrrr@{}}
    \toprule
    \textbf{Model} & \textbf{P@10} & \textbf{P@32} & \textbf{R@10} & \textbf{R@32} \\
    \midrule
    T5      & 8.685 & 15.26 & 2.672 & 4.624 \\
    SnapLGR        & \textbf{21.78} & \textbf{34.67} & \textbf{6.437} & \textbf{11.11} \\
    \midrule
    \textit{Ratio (SnapLGR / T5)} & \textit{$2.51\times$} & \textit{$2.27\times$} & \textit{$2.41\times$} & \textit{$2.40\times$} \\
    \bottomrule
  \end{tabular}
\end{table}

\subsubsection{Hypothesis 1: Impact of Model Architecture} 
\begin{figure}[t]
  \centering
  \includegraphics[width=\columnwidth]{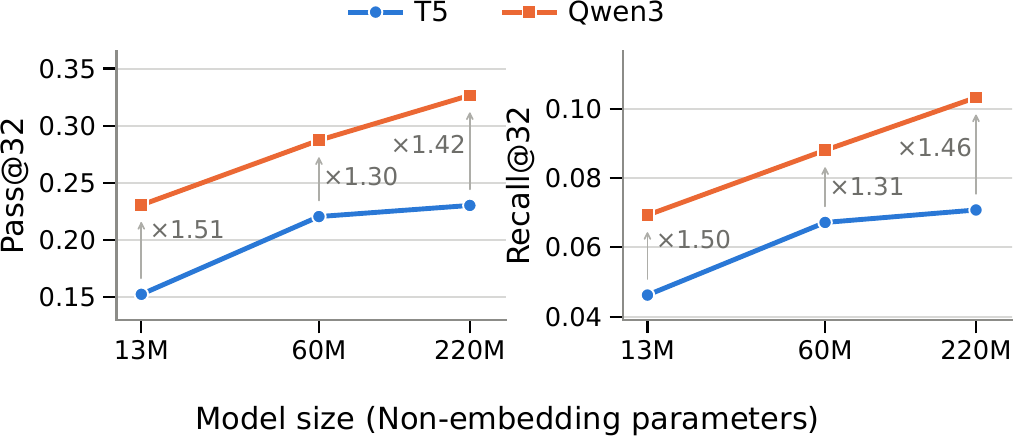}
\caption{Pass@32 (left) and Recall@32 (right) for Qwen3 and T5 architectures across parameter scales from 13M to 600M. Annotations indicate the relative performance multiplier of Qwen3 compared to T5 at matching scales. All models are trained using SFT only from random initialization.}
  \label{fig:size}
\end{figure}

Figure~\ref{fig:size} compares retrieval performance across randomly initialized decoder-only Qwen3 models and matched T5 encoder--decoder baselines. Note that all experiments use SFT only (no CPT) with random initialization to cleanly isolate structural efficiency. The Qwen3 decoder-only models consistently outperform the T5 baseline on both Pass@32 and Recall@32 at every matched parameter scale. Specifically, Qwen3 delivers a $1.51\times$ lift in Pass@32 and a $1.50\times$ lift in Recall@32 at the 13M scale, maintaining strong relative advantages ranging from $1.30\times$ to $1.46\times$ at the 60M and 220M scales. {This is the largest effect we isolate, and it does not appear substitutable by capacity alone: T5 at 220M reaches a Pass@32 of $23.03\%$, close to what Qwen3 attains at 13M ($23.07\%$).} This suggests that the decoder-only architecture is inherently better suited for generative retrieval tasks.

\subsubsection{Hypothesis 2: Effect of Model Scaling} Using the same experimental setup from Figure~\ref{fig:size}, we also observe the effect of raw parameter scaling. Both architectures exhibit strong positive scaling behavior, with absolute retrieval performance steadily climbing as the parameter count increases from 13M to 600M. {The two families appear to saturate at different points, however: T5 gains $+44.6\%$ in Pass@32 from 13M to 60M but only $+4.4\%$ from 60M to 220M, whereas Qwen3 still gains $+14.3\%$ over that interval.} Crucially, the structural advantage of the decoder-only architecture holds robustly across these different model capacities.

\begin{table}[t]
  \centering
  \caption{Benefit of LLM pretraining and grounding. All experiments use Qwen3-0.6B, with either Random or Pretrained initialization, and Qwen3-VL SIDs. 
  }
  \label{tab:cpt_ablation}
  \setlength{\tabcolsep}{5pt}
  \begin{tabular}{@{}lrrrr@{}}
    \toprule
    \textbf{Model} & \textbf{P@10} & \textbf{P@32}  & \textbf{R@10} & \textbf{R@32} \\
    \midrule
   (1) Random + SFT           & 20.88          & 34.07          & 6.106          & 10.77 \\
   (2) Pretrained + SFT       & 21.60          & 34.46          & 6.341          & 10.98 \\
   (3) Pretrained + CPT + SFT & \textbf{21.78} & \textbf{34.67} & \textbf{6.437} & \textbf{11.11} \\
    \midrule
    \textit{Lift (2) $\rightarrow$ (3)} & \textit{+0.83\%} & \textit{+0.62\%} & \textit{+1.51\%} & \textit{+1.19\%} \\
    \bottomrule
  \end{tabular}
\end{table}

\subsubsection{H3: Benefit of Pretraining and Grounding} Table~\ref{tab:cpt_ablation} isolates the impact of LLM pretraining and continued pretraining (CPT) on downstream performance. Initializing with pretrained weights (\textit{Pretrained + SFT}) provides a clear baseline boost across all metrics compared to training from scratch (\textit{Random + SFT}), improving P@10 from 20.88\% to 21.60\% and R@10 from 6.106\% to 6.341\%. Furthermore, incorporating continued pretraining (\textit{Pretrained + CPT + SFT}) yields further gains, achieving the best performance across all metrics (21.78\% P@10 and 11.11\% R@32). While these gains represent the smallest of the three evaluated effects, they remain  consistent and practically relevant in the context of industrial-scale traffic.

\begin{table}[t]
  \centering
  \caption{SID-level and Video-level retrieval recall metrics for Qwen3-0.6 backbones trained with MM, MM+, and Qwen3-VL SIDs (SFT-only).}
  \label{tab:combined_retrieval_metrics}
  \setlength{\tabcolsep}{12pt}
  \begin{tabular}{@{}lcccc@{}}
    \toprule
    \multirow{2}{*}{\textbf{Tokenizer}} & \multicolumn{2}{c}{\textbf{SID Recall}} & \multicolumn{2}{c}{\textbf{Video Recall}} \\
    \cmidrule(lr){2-3} \cmidrule(l){4-5}
    & \textbf{@10} & \textbf{@32} & \textbf{@10} & \textbf{@32} \\
    \midrule
    MM       & 6.102 & 10.58 & 1.393 & 2.720 \\
    MM+      & 5.979 & 10.30 & 1.364 & 2.616 \\
    Qwen3-VL & \textbf{6.341} & \textbf{10.98} & \textbf{1.416} & \textbf{2.765} \\
    \bottomrule
  \end{tabular}
\end{table}

\subsubsection{SID Tokenizer Ablations} 
Table~\ref{tab:combined_retrieval_metrics} isolates the effect of SID tokenization within a single 600M LLM setup. At the SID level, Qwen3-VL SIDs outperform legacy MM and MM+ tokenizers across all metrics (e.g., SID-Recall@32 of 10.98\% vs.\ 10.58\% and 10.30\%, respectively), and this advantage carries through downstream materialization to video-level retrieval, where Qwen3-VL reaches a Video-Recall@32 of 2.765\% against MM (2.720\%) and MM+ (2.616\%). {We note that while Qwen3-VL achieves the best performance in both downstream retrieval and collision resistance (\S\ref{sec:sid_quality_results}), the overall correlation between collision metrics and retrieval quality is not perfect. Specifically, the legacy MM tokenizer outperforms MM+ in retrieval despite exhibiting worse collision resistance. This suggests that while strong distributional properties are crucial prerequisites for reliable materialization, they are not absolute predictors of end-to-end performance, and further study is needed to fully disentangle their interplay.}

\subsection{Training and Inference Efficiency (RQ3)}
\label{sec:inference_results}

To address \textbf{RQ3}, we evaluate the impact of our system optimizations on training and inference throughput. 

\begin{table}[t]
  \centering
  \caption{Compounding training optimizations leading to production deployment. All experiments use a single node with 8 A100 GPUs,  besides the last row which uses a single node of 8 H100 GPUs. Speedups are cumulative relative to the un-optimized baseline, measured in terms of samples/s/GPU.
  }
  \label{tab:inference_macro}
  \setlength{\tabcolsep}{18pt}
  \begin{tabular}{@{}lc@{}}
    \toprule
    \textbf{Setting}  & \textbf{Speedup} \\
    \midrule
    Baseline  & 1.00$\times$ \\
    $\quad +$ \texttt{torch.compile} & 1.29$\times$ \\
    $\quad +$ FA2 Variable length support  &  1.47$\times$  \\
     $\quad +$ {Optimized packed sequences} &  1.78$\times$  \\
    $\quad +$ H100 GPUs   & \textbf{3.63$\times$} \\
    \bottomrule
  \end{tabular}
\end{table}

\subsubsection{Training optimization.} \label{sec:training_exps}
Table~\ref{tab:inference_macro} details the cumulative impact of our training optimizations on system throughput. Starting from a standard baseline on an 8 $\times$ NVIDIA A100-80GB node, applying \\
\texttt{torch.compile} yields an initial $1.29\times$ speedup by reducing framework overhead and fusing GPU operations. We further optimize attention computation by integrating FlashAttention-2 with variable-length support, which raises the relative throughput by $1.47\times$. We leverage dynamic sequence packing to eliminate wasted computation on padding tokens; by subsequently optimizing the target packed sequence length to maximize hardware utilization, we push the A100 throughput to $1.78\times$. Finally, migrating this fully optimized software stack to a single node equipped with NVIDIA H100 GPUs provides a substantial hardware-accelerated boost, achieving a peak end-to-end training throughput of $3.63\times$ relative to the baseline. These compounding software and hardware optimizations are critical for enabling SnapLGR's frequent periodic retraining within strict production budgets.

\subsubsection{Inference optimization.} \label{sec:inference_exps}
Our initial profiling revealed that standard beam search performance was primarily constrained by framework overhead rather than GPU computation. Specifically, Python-based beam scoring, KV-cache forking, coordination, and scheduler overhead dominated execution time. On a multi-GPU setup with beam width 32, actual GPU decoding accounted for only 12--15\% of total wall-clock time.

\noindent\textbf{TensorRT-LLM.} To address the primary bottleneck, we migrated the Python-based beam search implementation to TensorRT-LLM's CUDA-backed beam search runtime. This eliminated much of the beam management overhead and resulted in a 9.3$\times$ throughput improvement on a single A100 GPU at beam width 32, while maintaining equivalent retrieval accuracy. 

\noindent\textbf{Decentralized worker-loop architecture.} We redesigned the production inference workflow from a centralized driver-dispatch architecture to a \emph{decentralized worker-loop} processing model to improve distributed inference efficiency. Each worker independently claims, loads, and processes its assigned data shards, effectively removing head-node fan-out bottlenecks. Running this worker-loop architecture on a 64$\times$A100 cluster achieved an additional 3.3$\times$ throughput improvement over the centralized architecture. 

\noindent\textbf{Asynchronous I/O and System Overlap.} After optimizing the framework and distribution architecture, generation speed was bottlenecked by data I/O and CPU-side execution. We further optimized the worker-loop by executing background row-group prefetching to hide read latencies, isolating Parquet writes to a dedicated asynchronous pool to hide storage latencies, and disabling the execution of unused metric encoding operations during the generation loop. Together, this combination of I/O and system overlap optimizations provided another significant gain, multiplying throughput by 1.49$\times$ relative to the un-optimized worker-loop baseline.

\noindent\textbf{Cumulative System Gains.} As detailed in Table~\ref{tab:async_io_macro}, integrating these optimizations yielded compounding improvements across the serving stack. End-to-end, migrating from the legacy Python-based centralized architecture to the fully optimized TensorRT-LLM worker-loop delivered an approximate 45.7$\times$ increase in cumulative per-GPU throughput on A100 GPUs.

\begin{table}[t]
  \centering
  \caption{Compounding inference throughput optimizations. Speedups are cumulative relative to the baseline centralized Python-based architecture on A100 GPUs, measured in terms of samples/s/GPU.}
  \label{tab:async_io_macro}
  \setlength{\tabcolsep}{4pt}
  \begin{tabular}{@{}lc@{}}
    \toprule
    \textbf{Optimization} & \textbf{Speedup} \\
    \midrule
    Baseline Centralized Architecture  & 1.00$\times$ \\
    $\quad +$ TensorRT-LLM CUDA-backed beam search & 9.3$\times$ \\
    $\quad +$ Decentralized worker-loop architecture & 30.7$\times$ \\
    $\quad +$ Asynchronous I/O and System Overlap & \textbf{45.7$\times$} \\
    \bottomrule
  \end{tabular}
\end{table}

\subsection{Online Production Impact (RQ4)}
\label{sec:online_results}

Table~\ref{tab:online_ab} reports a 7-day production A/B test of SnapLGR against the
legacy T5 baseline. SnapLGR achieved statistically significant positive gains across core feed engagement metrics, highlighted by a {+0.37\%} relative increase in View Time, a {+0.09\%} increase in Time Spent, a +0.18\% increase in Deep Sessions, and a +0.11\% increase in Deep Sessions Unique User.  At our
platform's scale these relative lifts amount to a substantial absolute gain in aggregate
engagement hours, and the Deep Session lifts indicate that the improved candidate quality
drives sustained, high-intent engagement and return visits rather than merely longer immediate viewing.

{We note that this production comparison changes the tokenizer, backbone, and serving stack simultaneously, so
the A/B measures the launched system as a whole; \S\ref{sec:ablations} isolates individual
components offline under a fixed tokenizer.}

\begin{table}[t]
  \centering
  \caption{Online A/B results: SnapLGR vs.\ legacy T5 baseline.}
  \label{tab:online_ab}
  \setlength{\tabcolsep}{10pt}
  \begin{tabular}{@{}lcc@{}}
    \toprule
    \textbf{Metric} & \textbf{Relative Change} & \textbf{p-value} \\
    \midrule
    View Time        & {+0.37\% $\pm$0.27\%} & 0.007 \\
    Time Spent       & {+0.09\% $\pm$0.09\%} & 0.048 \\
    Deep Sessions          & +0.18\% $\pm$0.16\% & 0.027 \\
    Deep Sessions Unique User   & +0.11\% $\pm$0.09\% & 0.017 \\
    \bottomrule
  \end{tabular}
\end{table}

\section{Related Work}
\label{sec:related}

\noindent\textbf{Generative retrieval and SIDs.}
DSI~\cite{tay2022transformer} and GENRE~\cite{decao2021autoregressive} pioneered mapping queries to document identifiers via autoregressive generation. TIGER~\cite{rajput2023recommender} brought this paradigm to recommendation, deriving hierarchical SIDs from RQ-VAE~\cite{lee2022autoregressive} codebooks and training a seq2seq Transformer to generate them; \citet{singh2024better} further showed these IDs generalize to ranking. Recently, SID construction has been extended to incorporate multimodal signals~\cite{zhu2025beyond} and collaborative contrastive objectives~\cite{wang2024learnable}. Our work extends this thread by applying Personalized PageRank (PPR) signals to supervise the collaborative objective. We also build upon the non-collaborative RQ-VAE design documented in \citet{ju2026semanticids}, utilizing full-codebook straight-through optimization and layer normalization to mitigate codebook collapse.

\noindent\textbf{LLM-based recommendation at scale.}
Early frameworks like P5~\cite{geng2022recommendation} cast recommendation as text generation, inspiring subsequent efforts to ground pretrained LLMs in collaborative semantics via embedding alignment~\cite{liao2024llara} and instruction tuning~\cite{zheng2024adapting}. Recently, industrial recommendation systems have rapidly adopted large generative models across sequential transducers, unified retrieve-and-rank architectures, and personalized ranking tasks~\cite{zhai2024actions,han2025mtgr,deng2025onerec,zhou2025onerectr,liu2025onerecthink,zhou2025openonerec,yang2026onereason,yang2025cobra,firooz2025360brew,yi2025recgpt,liu2026tokenminds,huang2025towards}. The closest industrial precursor to our modeling pipeline is PLUM~\cite{he2026plum}, which adapts LLMs for generative retrieval at YouTube using SID tokenization, continued pretraining, and task-specific fine-tuning. While our training methodology closely follows PLUM (\S\ref{sec:training}), our work specifically addresses the systems-level challenges of production deployment, introduces PPR-based co-engagement supervision, evaluates SID quality and embedding grounding, and studies the drivers of improvement against conventional TIGER-style GR.

\section{Conclusion}
\label{sec:conclusion}

We present the end-to-end design, production launch, and empirical evaluation of SnapLGR, an LLM-based GR system for a short video platform at Snapchat. By jointly co-designing multimodal PPR-supervised SIDs, two-stage training including vocabulary grounding, and distributed training and inference, our system achieved significant online lifts in View Time, Time Spent, Deep Sessions, and Deep Sessions Unique User over the incumbent TIGER-style baseline. {Our deployment highlights two practical takeaways for practitioners. 
First, deploying LLM-based GR at industrial scale  requires co-optimizing the serving infrastructure; advanced inference techniques, such as CUDA-backed beam search and decentralized worker architectures, are essential to making wide-beam generation practically viable. Second, our offline attribution reveals that transitioning to a decoder-only architecture is the primary driver of the observed retrieval improvements, while enhanced tokenization and vocabulary grounding provide smaller, yet vital, compounding benefits. Together, these findings underscore the need for a holistic approach when scaling modern GR systems. For future work, we plan to explore scaling to larger LLM backbones to leverage their more extensive pretrained world knowledge for GR. 
}

\bibliographystyle{ACM-Reference-Format}
\bibliography{main}

\appendix

\section*{Acknowledgements}

Author roles are as follows:
\begin{itemize}
    \item \textit{Core Contributors:} Liam Collins, Jiwen Ren
    \item \textit{Contributors:} Donald Loveland, Bhuvesh Kumar, Clark Mingxuan Ju, Xuan Guo, Mo Li, Alvin Hou, Yi Cui, Peng Yang, Jian Wang, Saud Afzal Shafi, Nga Than, Ruiming Lu
    \item  \textit{Project Lead:} Wenfeng Zhuo, Dongheng Li 
    \item \textit{Leadership:} Lili Zhang, Mingtao Zhang, Jinchao Ye, Jiacong He, Vincent Xue, Chunhui Zhu, Neil Shah
\end{itemize}
We would also like to acknowledge Pau Kung and Jiahui Shi for their work in developing the internal large-scale graph processing framework that enabled the PPR-based positive pair construction for collaborative supervision during SID training.

\section{Additional Implementation Details}\label{app:pipeline-details}

\subsection{SID Pipeline}
\label{app:qwen_sid_details}

We use the architecture and straight-through estimator from \citet{ju2026semanticids}; Section~\ref{sec:sids} gives the mathematical formulation. Codebooks are orthogonally initialized and L2-normalized. The implementation applies a softmax to the cosine scores to form the backward-pass surrogate in Equation~\eqref{eq:full_codebook_ste}.

We use the first 1024 dimensions of the ambient 4096-dimensional Qwen3-VL-Embedding-8B embedding. This is possible because the model is trained with Matryoshka representation learning~\cite{kusupati2022matryoshka}, meaning embedding prefixes are also valid embeddings.

\subsubsection{Personalized PageRank (PPR) Formulation}
\label{app:ppr_details}

To compute the co-engagement signal for SID construction (Section~\ref{sec:sids}), we rely on Personalized PageRank (PPR)~\cite{haveliwala2002topic}. For each anchor video $i$, we extract its two-hop neighborhood from the bipartite user--video engagement graph and form a row-normalized transition matrix $\mathbf P$, where $P_{ab}$ is the probability of moving uniformly from node $a$ to one of its neighbors. The PPR score vector $\boldsymbol\pi_i$ is the stationary solution to:
\begin{equation}
  \boldsymbol\pi_i=(1-\alpha)\mathbf e_i+\alpha\mathbf P^\top\boldsymbol\pi_i,
  \label{eq:ppr}
\end{equation}
where $\mathbf e_i$ places all mass on anchor $i$ and $\alpha$ is the transition probability. Equivalently, a walk starts at $i$ and repeatedly either returns to $i$ with probability $1-\alpha$ or moves to a random neighboring node with probability $\alpha$. This mechanism ensures the resulting similarity scores remain centered on the anchor's local neighborhood rather than drifting toward globally popular regions of the graph.
All graph processing is executed using an internal large-scale graph processing framework.

\subsection{LLM Training Pipeline}
\subsubsection{CPT Prompt.} \label{app:cpt}
Figure \ref{fig:cpt_examples} gives an illustrative CPT prompt.

\begin{figure}[t]
  \centering
  \small
  \fbox{%
  \begin{minipage}{0.92\columnwidth}
    \textbf{CPT Prompt:}\\
    \texttt{Video
    <|sid\_begin|><s\_a\_857><s\_b\_412><s\_c\_10><|sid\_end|> is about a person dressed as a clown sitting on a ledge. A person walks up to the clown and they hug. The video is in black and white.}\\[3pt]
  \end{minipage}}
  \caption{Illustrative CPT training prompt. The video corpus grounds SID tokens in textual semantics.}
  \label{fig:cpt_examples}
\end{figure}

\subsection{Inference Pipeline}

\subsubsection{Value-weighted materialization.} We maintain an SID-to-Video ID lookup table to enable fast materialization of videos given generated SIDs.
Our lookup table retains up to a fixed number of candidate stories per SID. Within each SID bucket, stories are ranked using a multi-objective linear value function that weighs downstream engagement signals:
\begin{equation}
    V(i) = \sum_{m \in \mathcal{M}} w_m \cdot S_m(i),
\end{equation}
where $S_m(i)$ represents the estimated rate for metric $m$ and $w_m$ is the corresponding weight that favors stronger engagement signals.

\section{Additional Experimental Details and Results}
\label{app:additional_results}

This appendix reports the complete SID-level metric grids for the offline
analyses in Section~\ref{sec:experiments}. All values are percentages. Unless
noted otherwise, the backbone retrieval model is the pretrained-initialized Qwen3-0.6B SFT-only
600k-step checkpoint and inference uses beam width 32. We also provide additional details of the metrics and report the results of new experiments in the following subsections.

\begin{table}[h]
  \centering \setlength{\tabcolsep}{5pt}
  \caption{Full performance comparison between the legacy T5 encoder-decoder baseline and the reference SnapLGR.} 
  \label{tab:t5-vs-llm-headline}
  \begin{tabular}{lrrr}
    \toprule
    \textbf{Metric} & \textbf{T5} & \textbf{SnapLGR} & \textbf{Ratio} \\
    \midrule
    Pass@1        &  1.711 &  4.857 & $2.84\times$ \\
    Pass@5        &  5.608 & 14.78 & $2.63\times$ \\
    Pass@10       &  8.685 & 21.78 & $2.51\times$ \\
    Pass@32       & 15.26 & 34.67 & $\mathbf{2.27\times}$ \\
    \midrule
    Pos-1-Pass@1  &  0.762 &  2.051 & $2.70\times$ \\
    Pos-1-Pass@5  &  2.317 &  6.057 & $2.61\times$ \\
    Pos-1-Pass@10 &  3.469 &  8.853 & $2.55\times$ \\
    Pos-1-Pass@32 &  5.778 & 14.41 & $2.49\times$ \\
    \midrule
    Recall@1      &  0.556 &  1.318 & $2.36\times$ \\
    Recall@5      &  1.749 &  4.195 & $2.40\times$ \\
    Recall@10     &  2.672 &  6.437 & $2.41\times$ \\
    Recall@32     &  4.624 & 11.11 & $2.40\times$ \\
    \bottomrule
  \end{tabular}
\end{table}

\begin{table}[htbp]
  \centering
  \caption{Complete SID-level metric grid for the SID-type comparison of
  Table~\ref{tab:combined_retrieval_metrics} (models trained with MM, MM+, and
  Qwen3-VL SIDs).}
  \label{tab:sid_type_full}
  \setlength{\tabcolsep}{10pt}
  \begin{tabular}{lrrr}
    \toprule
    \textbf{Metric} & \textbf{MM} & \textbf{MM+} & \textbf{Qwen3-VL} \\
    \midrule
    Pass@1        &  4.719 &  4.616 &  4.896 \\
    Pass@5        & 14.29 & 13.94 & 14.74 \\
    Pass@10       & 21.02 & 20.49 & 21.60 \\
    Pass@32       & 33.60 & 32.76 & 34.46 \\
    \midrule
    Pos-1-Pass@1  &  1.987 &  1.951 &  2.082 \\
    Pos-1-Pass@5  &  5.800 &  5.672 &  6.053 \\
    Pos-1-Pass@10 &  8.504 &  8.293 &  9.022 \\
    Pos-1-Pass@32 & 13.88 & 13.45 & 14.38 \\
    \midrule
    Recall@1      &  1.279 &  1.263 &  1.331 \\
    Recall@5      &  4.005 &  3.938 &  4.170 \\
    Recall@10     &  6.102 &  5.979 &  6.341 \\
    Recall@32     & 10.58 & 10.30 & 10.98 \\
    \bottomrule
  \end{tabular}
\end{table}

\begin{table}[t]
  \centering
  \caption{SFT (SID-only, from random initialization) retrieval eval across downscaled Qwen3
  sizes. Sizes are backbone (non-embedding) params.}
  \label{tab:downscaling-sidonly}
  \setlength{\tabcolsep}{12pt}
  \begin{tabular}{lccc}
    \toprule
    \textbf{Metric} & \textbf{13M} & \textbf{60M} & \textbf{220M} \\
    \midrule
    Pass@1        &  2.60 &  3.56 &  4.28 \\
    Pass@5        &  8.49 & 11.01 & 13.19 \\
    Pass@10       & 13.05 & 16.77 & 19.62 \\
    Pass@32       & 23.07 & 28.73 & 32.66 \\
    \midrule
    Pos-1-Pass@1  &  1.10 &  1.53 &  1.83 \\
    Pos-1-Pass@5  &  3.28 &  4.39 &  5.37 \\
    Pos-1-Pass@10 &  4.94 &  6.54 &  7.99 \\
    Pos-1-Pass@32 &  8.48 & 11.13 & 13.38 \\
    \midrule
    Recall@1      &  0.80 &  1.04 &  1.19 \\
    Recall@5      &  2.51 &  3.14 &  3.76 \\
    Recall@10     &  3.83 &  4.89 &  5.80 \\
    Recall@32     &  6.93 &  8.80 & 10.32 \\
    \bottomrule
  \end{tabular}
\end{table}

\subsection{SID Quality Metrics}
\label{app:sid_metrics}

We provide further details on the metrics we use for our collision resistance study in Section~\ref{sec:sid_quality_results}.
 The following metrics are defined over a matched item corpus $\mathcal{I}$:

\begin{itemize}[leftmargin=*]
    \item \textbf{Utilization:} The proportion of the theoretical SID code space actively occupied by videos, defined as $|\mathcal{Z}_{\mathrm{obs}}| / \min(|\mathcal{I}|, \prod_{\ell=1}^L K_\ell)$, where $\mathcal{Z}_{\mathrm{obs}}$ is the set of realized SIDs and $K_\ell$ is the codebook size at level $\ell$.
    \item \textbf{Uniqueness:} The percentage of videos in the corpus whose complete assigned SID contains exactly one video.
    \item \textbf{Top-1\% Coverage:} An interpretable snapshot of extreme head concentration, measured by sorting realized SIDs by bucket size (descending) and calculating the fraction of total corpus videos contained within the largest 1\% of SID buckets.
    \item \textbf{Gini Coefficient:} A holistic summary of the collision distribution shape. It is calculated as the area between the cumulative video coverage curve (Figure~\ref{fig:tail_concentration}) and a perfectly uniform diagonal. A lower Gini score indicates that videos are evenly balanced across assigned SIDs, whereas a higher score (approaching 1.0) reveals severe concentration into a few massive ``mega-collision'' codes.
\end{itemize}

\subsection{Retrieval Metrics}\label{app:metrics}
Let $\mathcal{T}_r = \operatorname{set}(\widetilde{\mathcal{F}}_r)$ be the ground-truth target SIDs for request $r$, and let $\widehat{\mathcal{T}}_{r,k}$ consist of the model's top-$k$ generated SIDs. {Recall@$k$} is defined as $|\mathcal{T}_r \cap \widehat{\mathcal{T}}_{r,k}| / |\mathcal{T}_r|$. {Pass@$k$} is 1 if $\mathcal{T}_r \cap \widehat{\mathcal{T}}_{r,k}$ is not empty and 0 otherwise, and {Pos-1-Pass@$k$} (short for Position-1-Pass@$k$) indicates whether $\widehat{\mathcal{T}}_{r,k}$ contains the immediate next ground-truth SID $\mathbf{z}_{r, b_r+1}$. Metrics are averaged uniformly across samples and given as percentages.

\subsection{Hyperparameter Tuning} 
\noindent\textbf{Model scaling learning-rate selection.}
In the model scaling experiment, learning rates for the T5 retrievers were initially estimated with Step's Law \cite{li2025predictablescale} anchored from the previously-tuned learning rate of $1e-3$ at 13M parameters, and then tuned to as small as $5\times 10^{-5}$, decreasing with model width; the Qwen3 runs reuse the resulting per-size values. 
The model size, tuned learning rate pairs are: (13M, $1\times 10^{-3}$), (60M, $3.4\times 10^{-4}$), (220M, $2.6\times 10^{-4}$).

\begin{table}[htbp]
  \centering
  \caption{Effect of training data scale. Below gives metrics across checkpoints of an
  SFT-only (no CPT) training run. Columns give training progress as a fraction of one epoch.}
  \label{tab:data_volume_full}
  \setlength{\tabcolsep}{7pt}
  \begin{tabular}{lrrrrr}
    \toprule
    \textbf{Metric} & \textbf{0.2} & \textbf{0.4} & \textbf{0.6} & \textbf{0.8} & \textbf{1.0} \\
    \midrule
    Tokens (B)    &   69.1 &  138.2 &  207.4 &  276.5 &  345.6 \\
    \midrule
    Pass@1        &   3.89 &   4.37 &   4.60 &   4.81 &   4.90 \\
    Pass@5        &  12.00 &  13.27 &  13.87 &  14.54 &  14.74 \\
    Pass@10       &  17.82 &  19.54 &  20.38 &  21.34 &  21.60 \\
    Pass@32       &  28.93 &  31.28 &  32.60 &  34.12 &  34.46 \\
    \midrule
    Pos-1-Pass@1  &   1.60 &   1.83 &   1.94 &   2.03 &   2.08 \\
    Pos-1-Pass@5  &   4.70 &   5.33 &   5.65 &   5.93 &   6.05 \\
    Pos-1-Pass@10 &   6.90 &   7.80 &   8.25 &   8.67 &   8.85 \\
    Pos-1-Pass@32 &  11.27 &  12.63 &  13.38 &  14.13 &  14.38 \\
    \midrule
    Recall@1      &   1.08 &   1.20 &   1.26 &   1.31 &   1.33 \\
    Recall@5      &   3.39 &   3.75 &   3.93 &   4.10 &   4.17 \\
    Recall@10     &   5.14 &   5.69 &   5.97 &   6.24 &   6.34 \\
    Recall@32     &   8.85 &   9.77 &  10.27 &  10.83 &  10.98 \\
    \bottomrule
  \end{tabular}
\end{table}

\subsection{Training Data Scaling}

Table~\ref{tab:data_volume_full}  evaluates five checkpoints from one pretrained-initialized Qwen3-0.6B SFT-only run; no CPT is used. Recall improves monotonically at every cutoff: from 0.2 to 1 epoch, Recall@32 rises from 8.85\% to 10.98\% ($+2.126$ percentage points, or $+24.0\%$ relative). The final 0.2-epoch interval adds only 0.149 points, suggesting that gains taper at the end of the observed range. The Pass@32 and Pos-1-Pass@32 metrics also rise, from 28.93\% to 34.46\% and from 11.27\% to 14.38\%, respectively.

\begin{table}[htbp]
  \centering
  \caption{Complete metrics for the initialization comparison of
  Table~\ref{tab:cpt_ablation}  with both runs SFT-only,  no CPT.
  Deltas are pretrained relative to random-init. 
  }
  \label{tab:init_full}
  \setlength{\tabcolsep}{5pt}
  \begin{tabular}{lccc}
    \toprule
    \textbf{Metric} & \textbf{Random} & \textbf{Pretrained} & \textbf{$\Delta$ (\%)} \\
    \midrule
    Pass@1        &  4.578 & \textbf{ 4.896} & $+6.9\%$ \\
    Pass@5        & 14.08 & \textbf{14.74} & $+4.7\%$ \\
    Pass@10       & 20.88 & \textbf{21.60} & $+3.4\%$ \\
    Pass@32       & 34.07 & \textbf{34.46} & $+1.1\%$ \\
    \midrule
    Pos-1-Pass@1  &  1.936 & \textbf{ 2.082} & $+7.5\%$ \\
    Pos-1-Pass@5  &  5.724 & \textbf{ 6.053} & $+5.7\%$ \\
    Pos-1-Pass@10 &  8.441 & \textbf{ 8.847} & $+4.8\%$ \\
    Pos-1-Pass@32 & 14.00 & \textbf{14.38} & $+2.7\%$ \\
    \midrule
    Recall@1      &  1.259 & \textbf{ 1.331} & $+5.7\%$ \\
    Recall@5      &  3.986 & \textbf{ 4.170} & $+4.6\%$ \\
    Recall@10     &  6.106 & \textbf{ 6.341} & $+3.8\%$ \\
    Recall@32     & 10.77 & \textbf{10.98} & $+1.9\%$ \\
    \bottomrule
  \end{tabular}
\end{table}

\begin{table}[htbp]
  \centering
  \caption{Effect of beam width on SID-level metrics for baseline SFT-only checkpoint.}
  \label{tab:beam_width}
  \setlength{\tabcolsep}{3pt}
  \begin{tabular}{lrrrr}
    \toprule
    \textbf{Metric} & \textbf{Beam 32} & \textbf{Beam 96} & \textbf{$\Delta$ (pp)} & \textbf{$\Delta$ (\%)} \\
    \midrule
    Pass@1        &  4.90 &  4.90 & $+$0.008 &  $+$0.16 \\
    Pass@5        & 14.74 & 14.85 & $+$0.113 &  $+$0.77 \\
    Pass@10       & 21.60 & 22.02 & $+$0.416 &  $+$1.93 \\
    Pass@32       & 34.46 & 37.90 & $+$3.448 & $+$10.01 \\
    \midrule
    Pos-1-Pass@1  &  2.08 &  2.09 & $+$0.003 &  $+$0.13 \\
    Pos-1-Pass@5  &  6.05 &  6.10 & $+$0.045 &  $+$0.74 \\
    Pos-1-Pass@10 &  8.85 &  9.02 & $+$0.173 &  $+$1.96 \\
    Pos-1-Pass@32 & 14.38 & 16.03 & $+$1.654 & $+$11.50 \\
    \midrule
    Recall@1      &  1.33 &  1.33 & $-$0.000 &  $+$0.07 \\
    Recall@5      &  4.17 &  4.19 & $+$0.022 &  $+$0.53 \\
    Recall@10     &  6.34 &  6.44 & $+$0.101 &  $+$1.59 \\
    Recall@32     & 10.98 & 12.25 & $+$1.269 & $+$11.56 \\
    \bottomrule
  \end{tabular}
\end{table}

\subsection{Beam Width Latency-Recall Trade-off.} 
Table  \ref{tab:beam_width} shows that increasing the beam width from 32 to 96 leads to substantial gains for metrics with larger $k$, for example Recall@$32$ increased by 11.56\%.
However, moving from width 32 to 96 reduced throughput by 45.0\% on A100 GPUs.

\subsection{SID Quality Case Study}
\label{app:sid_qualitative}

\begin{figure}[t]
  \centering
  \includegraphics[width=\columnwidth]{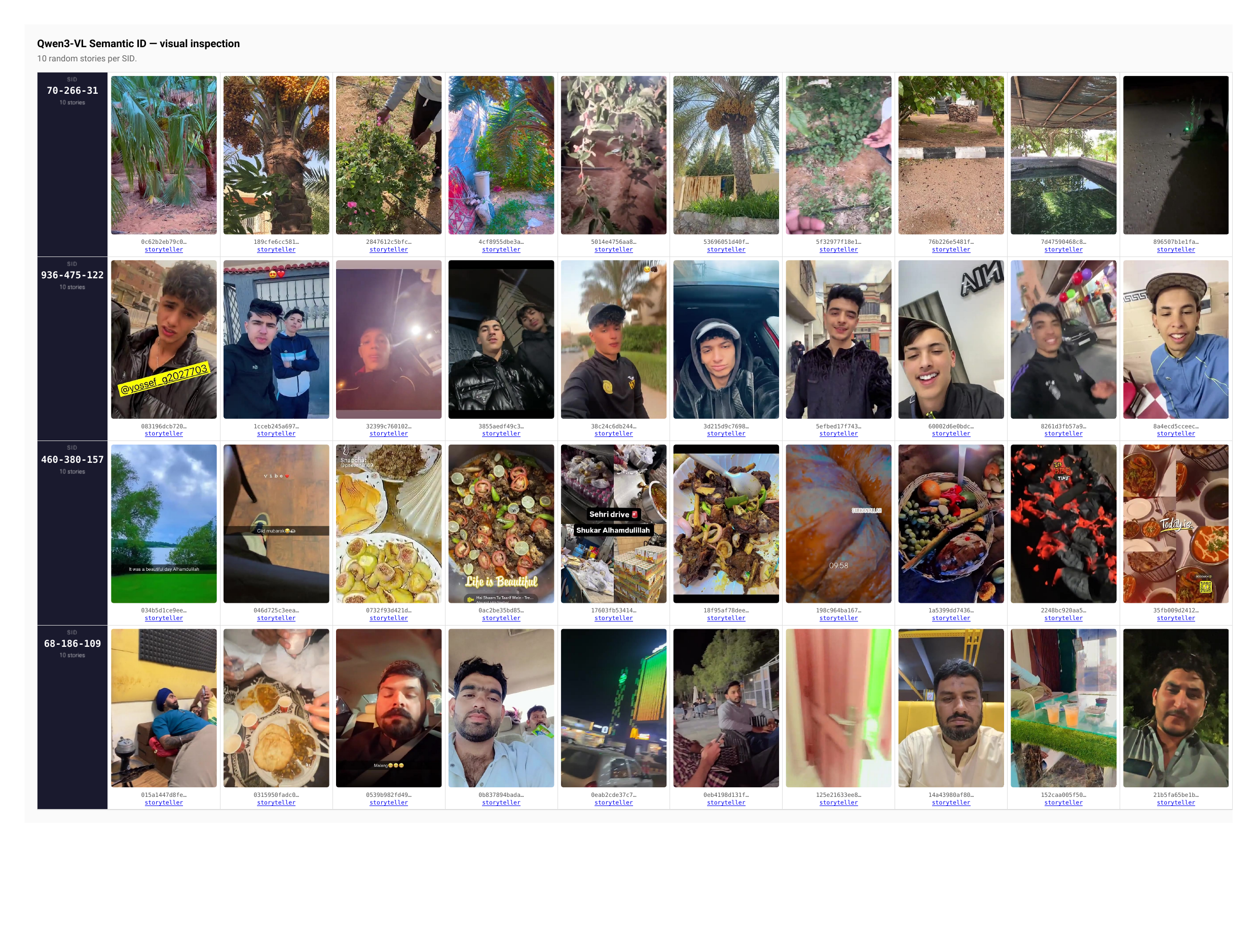}
  \caption{Example SID-to-video mappings. Each row contains images from 10 randomly-selected videos with the same SID. Videos with the same SID are highly semantically-similar.}
  \label{fig:sid-vis}
\end{figure}

To qualitatively assess the semantic fidelity of our Qwen3-VL-based multimodal and co-engagement-supervised tokenization pipeline, we visually inspect videos that are mapped to the same Semantic ID (SID). Figure~\ref{fig:sid-vis} illustrates several example SID-to-video mappings, where each row displays keyframes from 10 randomly selected videos sharing an identical complete SID. As demonstrated in the figure, videos assigned to the same SID exhibit strong visual and topical coherence. This visual consistency confirms that the multimodal, behavior-infused RQ-VAE successfully preserves content semantics while compressing dense item embeddings into discrete identifier buckets.

\end{document}